\documentclass[journal]{IEEEtran}
\ifCLASSINFOpdf
\else
\fi
\usepackage{color}
\usepackage{graphicx}
\usepackage{amsmath}
\usepackage{tabularx}
\usepackage{multirow}
\usepackage{booktabs}
\usepackage{cite}
\usepackage{datetime}
\usepackage{amssymb}
\usepackage{float}
\usepackage[T1]{fontenc}

\begin{document}

%
\title{Management of Cascading Outage Risk Based on \\Risk Gradient and Markovian Tree Search}
%
%
%

\author{Rui~Yao,~\IEEEmembership{Member,~IEEE,}~Kai~Sun,~\IEEEmembership{Senior~Member,~IEEE,}~Feng~Liu,~\IEEEmembership{Member,~IEEE,}~Shengwei~Mei,~\IEEEmembership{Fellow,~IEEE}
\thanks{This work was supported by the CURENT Engineering Research Center.}
\thanks{R. Yao and K. Sun are with the Department of EECS, University of Tennessee, Knoxville, TN 37996, USA (emails: yaorui.thu@gmail.com, kaisun@utk.edu).}
\thanks{F. Liu and S. Mei are with the State Key Laboratory of Power Systems, Department of Electrical Engineering, Tsinghua University, Beijing 100084, China. (email: lfeng@mail.tsinghua.edu.cn)}}

%
%

\markboth{}%
{Shell \MakeLowercase{\textit{et al.}}: Bare Demo of IEEEtran.cls for IEEE Journals}
%



\maketitle

\begin{abstract}
Since cascading outages are major threats to power systems, it is important to reduce the risk of potential cascading outages. In this paper, a risk management method of cascading outages based on Markovian tree search is proposed. With the tree expansion on the cascading outage risk, risk gradient is computed efficiently by a forward-backward tree search scheme with good convergence, and it is then employed in an optimization model to minimize control cost while effectively reducing the cascading outage risk. To overcome the limitation with linearization in computing risk gradient, an iterative risk management (IRM) approach is further developed. Tests on the RTS-96 3-area system verify the accuracy of the computed risk gradient and its effectiveness for risk reduction. Time performance of the proposed IRM approach is tested on the RTS-96 system, a 410-bus US-Canada northeast system and a 1354-bus Mid-European system, and demonstrates its potentials for decision support on practical power systems online or on hourly basis.
\end{abstract}

\begin{IEEEkeywords}
cascading outage, tree search, risk management, risk gradient, control cost, trade-off, multi-objective optimization
\end{IEEEkeywords}

%
\IEEEpeerreviewmaketitle

\section{Introduction}

\IEEEPARstart{C}{ascading} outages are major threats to power system operations. To study and prevent cascading outages, people have proposed various methods on the simulation, risk assessment and risk management of cascading outages\cite{hines2009large}.
The mitigation methods\cite{carreras2003blackout,chen2005cascading,newman2011exploring} based on power-law analysis can provide rough mitigation strategies in the planning horizon, but lack accurate tactics for online operations. In grid operations, an extensively used approach is to check deterministic N-1 or N-k criteria to guarantee no limit being violated under given contingencies. Another methodology is robust optimization\cite{bertsimas2013adaptive}, which ensures the strategy to cover all foreseen high-risk scenarios. However, the prevention of all potential cascading outages could be too expensive to achieve for both planning and operations of the system. Also it is practically infeasible to completely wipe out the risk of any cascading outages\cite{vaiman2012risk}. Although the deterministic methods are more computationally efficient, they cannot distinguish the actual probabilities and consequences of different events, so the control cost could be unnecessarily spent on preventing some low-risk events\cite{kirschen2007comparison}. Therefore, it is more desirable to introduce the index of risk (i.e. the expectation of consequences) to provide more effective guidance for decision-making in power system operations.

Several risk-based approaches have been proposed in the recent years. Some methods apply risk indices based on pre-defined severity functions to optimization models \cite{xiao2007risk,wang2013computational,wang2013risk,wang2014risk}. The risk indices in these methods quantify the expected violation of operation constraints as indirect representations of the risk faced by the end users \cite{kirschen2007comparison}. While some methods establish the risk management under a given contingency set as a chance-constrained optimization problem \cite{karangelos2016probabilistic,hamon2015computational}. Also some methods incorporate reliability indices into the economic dispatch or unit commitment models\cite{simopoulos2006reliability,ortega2016assessment,capitanescu2015enhanced}, and realizes coordination between the economic profit and operation risk\cite{he2010optimising}. However, these methods did not consider the dependency among outages and hence may underestimate the risk when assessing cascading outages. Moreover, many efficient reliability assessment and risk management methods for independent contingencies are no longer effective when dealing with dependent outages. In summary, there is not yet a method designed for the risk management against dependent cascading outages. Such risk management should be able to provide direct risk measurement in the amount of load shed, energy shortage or economic loss. Moreover, to avoid omitting unaware risky scenarios, the effective risk management should not be confined in a pre-set contingency set\cite{kirschen2007comparison}. It is rather, desirable to efficiently search the most risky set of cascading outage patterns for risk management. 

To realize effective and efficient risk management, all the following features are desirable:

1) Reasonable simulation and efficient risk assessment of multi-timescale dependent cascading outages: firstly, a reasonable modeling and simulation method should reflect the primary characteristics of cascading outages, e.g. the dependency among outages, timescales of related processes, etc. \cite{yao2014cascading,vaiman2012risk}; then, from numerous simulated outages, a risk assessment method needs to identify the most risky cascade paths; finally, the assessed risk is used in a risk management method to reduce the risks of cascading outages. 

2) Risk metrics that directly evaluate the risk faced by the end users for effective risk management, such as load or energy loss, and economic loss.

3) A risk management formulation allowing direct adjustment on reduction of risk, e.g. the amount of desired decrease in the load loss, energy loss or economic loss, which will facilitate to determining the trade-off between risk reduction and control cost.

4) An efficient risk management algorithm to derive strategies for risk reduction on cascading outages.

Note that failing to realize any of the above desirable features would downgrade the effectiveness and efficiency in risk management of cascading outages. This paper proposes a risk management method having all the above features. To meet the feature 1, the quasi-dynamic simulation method in \cite{7254205} and the Markovian tree (MT) based risk assessment method in \cite{yao2016risk} are applied in this paper. The former can better reflect the multi-timescale nature and dependencies in cascading outages, and latter can efficiently identify the most risky cascading outage patterns and provide a variety of practical risk metrics. To meet the features 2 and 3, this paper introduces risk gradient and risk constraint, which allow direct adjustment on how much risk to be reduced on load loss, energy shortage and economic loss. Moreover, to overcome the limitation of linearization in the risk gradient computation, an iterative risk management (IRM) approach is proposed for more effective reduction of risk to meet feature 4. Although risk gradient can be calculated in multiple ways, e.g. the cumbersome perturbation method, this paper proposes an efficient forward-backward algorithm to calculate risk gradient using the MT in order for seamless integration into the MT search algorithm for risk assessment. Thus, risk assessment and risk gradient computation are performed at the same time to improve the overall efficiency of the proposed approach.

The rest of the paper is organized as follows. Section II first briefly introduces our previous work selected as a foundation of the proposed approach (Section II.A). With the MT expansion of risk, the generic model of risk management is proposed based on the concept of risk gradient (Section II.B). 
Section III is the analytical derivation of the risk gradient and the efficient algorithm for calculating the risk gradient. The risk gradient is derived in Section III.A-C. The computation of risk gradient is realized as an efficient forward-backward tree-search algorithm, which is integrated with the MT-based risk assessment algorithm (III.D).
After deriving the risk gradient, the Section IV implements the risk management based on the generic form introduced in Section II: Section IV.A proposes the full optimization model of risk management (RM); Section IV.B further presents the iterative risk management (IRM) model to overcome the limitation of linearization; and then Section IV.C discusses the application potentials of the proposed approach.
Section V is the test cases. Section V.A verifies the accuracy and the efficiency of the risk gradient computation on the RTS-96 test system. And then the RM and IRM are tested on RTS-96 test system, a US-Canada 410-bus system, and a 1354-bus Mid-European system, respectively.

\section{Generic Model of Risk Management Based on MT and Risk Gradient}

\subsection{Retro of risk assessment using Markovian tree search}
\begin{figure}[htb]
  \centering
  \includegraphics[clip=true,scale=0.2]{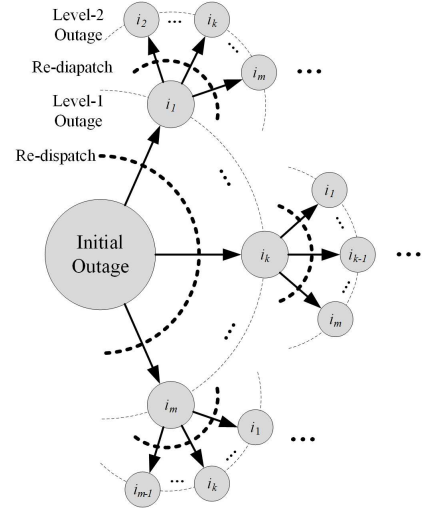}\\
  \caption{MT representation of cascading outage paths}\label{fig:MT_represent}
\end{figure}

The risk management needs quantification of risk. This paper selects our previous work on efficient risk assessment method with Markovian tree search \cite{yao2016risk}, and the quasi-dynamic simulation method \cite{7254205}. All the possible cascading outage paths are organized in a MT, as shown in Fig. \ref{fig:MT_represent}.

The time elapse is incorporated in the MT model, where each level on the MT corresponds to a time interval $\tau_D$, and each node is labeled with the outage sequence from the root, as $(i_{k_1}\cdots i_{k_n})$, where $i_{k_r}$ is either a positive integer denoting the index of the element failed on the $r$th level or 0 if no outage occurs on this level, $n$ is the ending level of the cascading outage. The cost corresponding to state $(i_{k_1}\cdots i_{k_n})$ is $C(i_{k_1} \cdots i_{k_n})$, and the conditional probability of outage event $i_{k_{r+1}}$ after the state $(i_{k_1} \cdots i_{k_r})$ is $\mathrm{Pr}(i_{k_{r+1}}|i_{k_1} \cdots i_{k_r})$. Then with the MT structure, the risk is expressed as the following expansion \cite{yao2016risk}:

\begin{equation}\label{eqn:MT_expansion}
\begin{aligned}
& \resizebox{1.0\hsize}{!}{$R=C_0+\displaystyle\sum_{k_1}\mathrm{Pr}(i_{k_1})C(i_{k_1})+\displaystyle\sum_{k_1}\mathrm{Pr}(i_{k_1})\displaystyle\sum_{k_2}\mathrm{Pr}(i_{k_2}|i_{k_1})C(i_{k_1}i_{k_2})$}\\
& \resizebox{1.0\hsize}{!}{$+\displaystyle\sum_{k_1}\mathrm{Pr}(i_{k_1})\displaystyle\sum_{k_2}\mathrm{Pr}(i_{k_2}|i_{k_1})\displaystyle\sum_{k_3}\mathrm{Pr}(i_{k_3}|i_{k_1}i_{k_2})C(i_{k_1}i_{k_2}i_{k_3})+\cdots$}
\end{aligned}
\end{equation}

Suppose control actions are taken to reduce the cascading outage risk after the initial outages. Then the expansion (\ref{eqn:MT_expansion}) can be divided into two parts: $C_0$ is the cost of the control, and all the other terms are the risks of subsequent cascading outages, whose sum is denoted by $R'$. The risk is assessed by searching the MT and adding the risk terms into (\ref{eqn:MT_expansion}) corresponding to the newly-visited states. 
To further accelerate the risk assessment, a forward-backward tree search algorithm based on a risk estimation index (REI) was proposed in \cite{yao2016risk}.
%

\subsection{Generic optimization model of risk management}
Cascading outages in the early stage usually develop slowly, so there is some time to adjust system states to reduce the risk of potential cascading outages after the initial outage. 
Control measures after the initial outages result in cost $C_0$ while reducing the subsequent risk $R'$, so a compromise between the effect and cost of risk management should be concerned. It is desirable that the risk of subsequent cascading outages is confined below a certain level $R'_{S}$, while the cost of control is minimized. Therefore, the basic formulation of risk management can be written as
\begin{equation}\label{eqn:mitigation_prototype}
\begin{aligned}
\min{}&f=C_0(\textbf{x}^*)\\
s.t.\ &R'(\textbf{x}^*)\leq R'_{S}\\
& \textbf{g}(\textbf{x}^*)\leq \textbf{0}
\end{aligned}
\end{equation}

\noindent where $\textbf{x}^*$ is the target system state, $C_0(\textbf{x}^*)$ is the cost of control, and $R'(\textbf{x}^*)$ is the subsequent risk. The last constraint represents the constraints in operations, e.g. load, generation and transmission capacity constraints. In real power systems, the commonly used control measures for overloading relief or power balance include re-dispatch of generation and load shedding initiated by operators or emergency control facilities. Moreover, to prevent voltage instability, reactive power compensation devices can also be adjusted. In emergency, the under-voltage load shedding, under-frequency load shedding and controlled islanding can also be triggered to prevent system-wide collapse. In this paper, cascading outages are modeled and simulated with DC power flow model, and the the re-dispatch operations for overloading relief are considered, which includes the re-dispatch of generators and curtailment of loads controlled by the operators. The re-dispatch is very commonly used in the operations of power systems, and it can be described with an optimal power flow model. The cost of re-dispatch $C_0(\textbf{x}^*)$ can be described as follows:

\begin{equation}\label{eqn:controlcost}
C_0(\textbf{x}^*)
=-\textbf{c}_D^{\mathrm{T}}(\textbf{P}^*_d-\textbf{P}_d)
+\textbf{c}_G^{\mathrm{T}}|\textbf{P}^*_g-\textbf{P}_g|
\end{equation}

Here $\textbf{x}^*=[\textbf{P}^*_d{}^{\mathrm{T}},\textbf{P}^*_g{}^{\mathrm{T}}]^{\mathrm{T}}$ is a vector of target re-dispatch state. $\textbf{x}=[\textbf{P}_d^{\mathrm{T}},\textbf{P}_g^{\mathrm{T}}]^{\mathrm{T}}$ is the pre-control system state. $\textbf{c}_D$ and $\textbf{c}_G$ are per unit costs of load shedding and generation adjustment, respectively. It should be noted that the proposed risk management approach is not limited to the re-dispatch of active power of generators and loads. For example, since the simulation and risk assessment can also be implemented in an AC power flow model, in that case the control of reactive power compensators can also be realized.

However, quantifying the $R'$ as a function of $\textbf{x}^*$ in (\ref{eqn:mitigation_prototype}) is not straightforward. Since cascading outages involve complex dependent events, any change in the system state will affect all the following states, and thus the risk terms on all the levels of the MT are changed. Therefore, it is infeasible to analytically quantify $R'$ as a function of $\textbf{x}^*$, but the risk at the original target state $\textbf{x}^*_0$ can be linearized to obtain the risk gradient:
\begin{equation}\label{eqn:gradient_prototype}
{\bf \Gamma}=
\left.\frac{\partial R'}{\partial \textbf{x}^*}\right|_{\textbf{x}^*=\textbf{x}^*_0}
\end{equation}
\noindent then the risk management model (\ref{eqn:mitigation_prototype}) becomes
\begin{equation}\label{eqn:mitigation_linearization}
\begin{aligned}
\min{}&f=C_0(\textbf{x}^*)\\
s.t.\ &{\bf \Gamma}\cdot(\textbf{x}^*-\textbf{x}^*_0)\leq R'_{S}-R'_{0}\\
& \textbf{g}(\textbf{x}^*)\leq \textbf{0}
\end{aligned}
\end{equation}

The next section will address the calculation of the risk gradient by using the result of risk assessment.

\section{Calculation of risk gradient}
\subsection{Derivative chain of states on the MT}
\begin{figure}[htb]
  \centering
  \includegraphics[clip=true,scale=0.09]{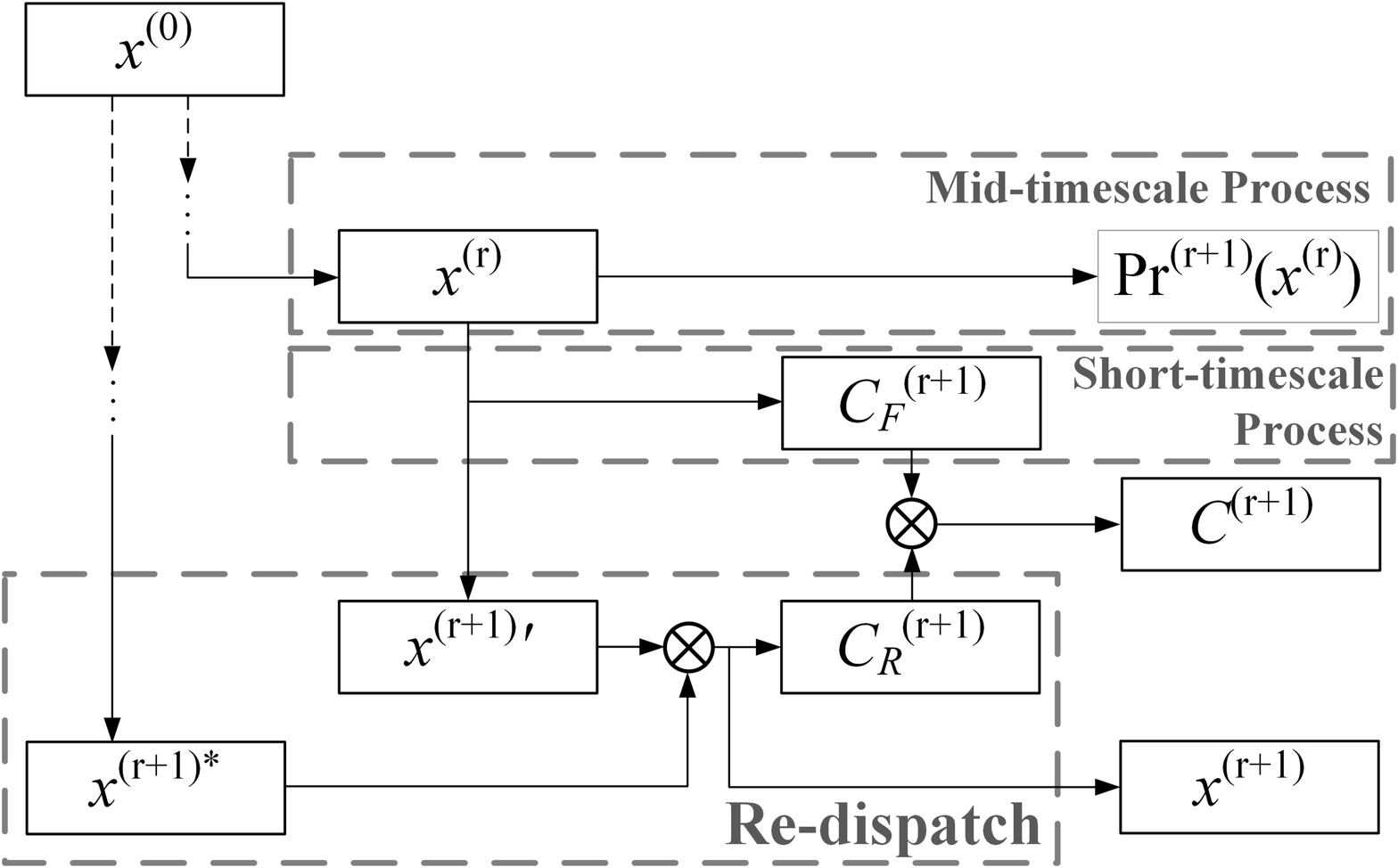}\\
  \caption{State dependencies on one level of cascading outage}\label{fig:state_chain}
\end{figure}

From (\ref{eqn:MT_expansion}) and (\ref{eqn:mitigation_linearization}), the risk gradient depends on the derivatives of conditional probability $\mathrm{Pr}(i_{k_r}|i_{k_1}\cdots i_{k_{r-1}})$ and cost $C(i_{k_1}\cdots i_{k_r})$ of states on each level of the MT. Such calculation requires the analysis of the chain of states on the cascading outage path. As shown in Fig.\ref{fig:state_chain}, denote the post-outage state on the $r$th level as $\textbf{x}^{(r)}{}'$, and the state after re-dispatch as $\textbf{x}^{(r)*}$. The costs of the short-timescale process and re-dispatch of all $i_{k_r} (r=1,\cdots,n)$ are briefly denoted as vectors $\textbf{C}_F^{(r)}$ and $\textbf{C}_R^{(r)}$, respectively, and the total cost on level $r$ is $\textbf{C}^{(r)}=\textbf{C}_F^{(r)}+\textbf{C}_R^{(r)}$. Briefly denote outage probabilities as a vector $\textbf{Pr}^{(r)}$.

In this paper, assume that the probabilities and the costs are differentiable. If in reality these quantities are non-smooth or discontinuous functions of states, then they need to be treated in segments, or the sub-derivatives are used as an approximation. And if there are discrete variables, then they need to be treated as continuous ones temporarily. 
To calculate the gradient of risk, the terms $\frac{\partial \textbf{Pr}^{(r)}}{\partial \textbf{x}^{(0)}}$ and $\frac{\partial \textbf{C}^{(r)}}{\partial \textbf{x}^{(0)}}$ are necessary according to (\ref{eqn:MT_expansion}). In fact, according to Fig.3, if the partial derivatives on levels up to $r$ (i.e. $\frac{\partial \textbf{C}^{(r)}}{\partial \textbf{x}^{(r-1)}}$, $\frac{\partial \textbf{Pr}^{(r)}}{\partial \textbf{x}^{(r-1)}}$ and $\frac{\partial \textbf{x}^{(r)}}{\partial \textbf{x}^{(r-1)}}$) are obtained, then $\frac{\partial \textbf{Pr}^{(r)}}{\partial \textbf{x}^{(0)}}$ and $\frac{\partial \textbf{C}^{(r)}}{\partial \textbf{x}^{(0)}}$ can be calculated iteratively. The derivation of the partial derivatives on each level depends on the analysis of the cascading process, which will be elucidated below.

\subsubsection{Mid-term Random Outage}

The probability of element $i$ outage on the MT is \cite{yao2016risk}:
\begin{equation}\label{eqn:prob_MT}
\mathrm{Pr}_i^{\mathrm{MT}}=
\frac{\lambda_i}{\sum_{j}\lambda_{j}}\left(1-\mathrm{e}^{-\sum_{j}\lambda_j\tau_D}\right)
\end{equation}

\noindent where $\lambda_i$ is the failure rate of branch $i$, which is assumed to be a function of its branch flow $F_i$ \cite{dobson2001initial,yang2016interval,chen2005cascading}. And $F_i$ is a function of the system state $\textbf{x}$. So the partial derivative of branch outage probability on level $r+1$ to $\textbf{x}^{(r)}$ is

\begin{equation}\label{eqn:part_prob_state}
\frac{\partial\textbf{Pr}^{(r+1)}}{\partial \textbf{x}^{(r)}}=
\frac{\partial\textbf{Pr}^{(r+1)}}{\partial {\bf \lambda}}\cdot
\frac{\partial{\bf \lambda}}{\partial \textbf{F}}\cdot
\left[-\textbf{y}_D\textbf{MY}^+,\textbf{y}_D\textbf{MY}^+\right]
\end{equation}
\noindent where ${\bf \lambda}$ and $\textbf{F}$ are vectors of $\lambda_i$ and $F_i$, respectively. $\textbf{y}_D$ is a diagonal matrix of branch admittances. $\textbf{M}$ is a $|V|\times|E|$ matrix and each of its column $M_{i_k}$ corresponding to a branch $i_k=\{u,v\}$ satisfies $M_{i_k,u}=1$, $M_{i_k,v}=-1$ and all the other entries are 0. $\textbf{Y}^+$ is the Penrose-Moore pseudo-inverse of admittance matrix $\textbf{Y}$.

\subsubsection{Short-timescale process}

\begin{figure}[htb]
  \centering
  \includegraphics[clip=true,scale=0.12]{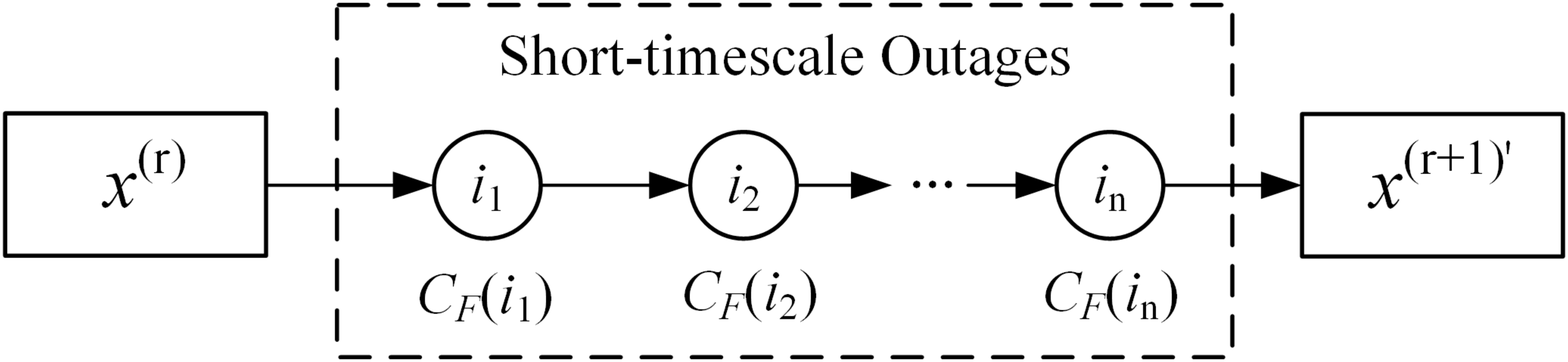}\\
  \caption{Illustration of short-timescale outage events and costs}\label{fig:short_chain}
\end{figure}

As Fig. \ref{fig:short_chain} shows, a short-timescale process may comprise of several outages, and each outage may directly lead to cost on the loss of load due to load shedding\cite{ren2008long}. The cost of short-timescale outages on level $r+1$ is the sum of costs caused by all outages.

\begin{equation}\label{eqn:cost_shorttimescale}
\textbf{C}_F^{(r+1)}=\sum_{k=1}^{n_{r+1}}\textbf{C}_F^{(k|r+1)}
\end{equation}

\noindent where $n_{r+1}$ is the number of short-timescale outages on level $r+1$, and the cost of the $k$th outage is $\textbf{C}_F^{(k|r+1)}$. $\textbf{x}^{(k|r+1)}$ is the state after the $k$th outage on level $r+1$ (note that $\textbf{x}^{(0|r+1)}=\textbf{x}^{(r)}$ and $\textbf{x}^{(n_{r+1}|r+1)}=\textbf{x}^{(r+1)}{}'$). Also $\frac{\partial \textbf{x}^{(k|r+1)}}{\partial \textbf{x}^{(k-1|r+1)}}$ and $\frac{\partial \textbf{C}_F^{(k|r+1)}}{\partial \textbf{x}^{(k-1|r+1)}}$ can be derived by sensitivity analysis of load shedding\cite{reddy2014sensitivity,girgis2010application}.
As the simulation of each outage on level $r+1$ continues, $\frac{\partial \textbf{x}^{(k|r+1)}}{\partial \textbf{x}^{(r)}}$ can be derived with:

\begin{equation}\label{eqn:deriv_short_state}
\frac{\partial \textbf{x}^{(k|r+1)}}{\partial \textbf{x}^{(r)}}=
\frac{\partial \textbf{x}^{(k|r+1)}}{\partial \textbf{x}^{(k-1|r+1)}}
\frac{\partial \textbf{x}^{(k-1|r+1)}}{\partial \textbf{x}^{(r)}}
\end{equation}

\begin{equation}\label{eqn:deriv_short_cost}
\frac{\partial \textbf{C}_F^{(k|r+1)}}{\partial \textbf{x}^{(r)}}=
\frac{\partial \textbf{C}_F^{(k|r+1)}}{\partial \textbf{x}^{(k-1|r+1)}}
\frac{\partial \textbf{x}^{(k-1|r+1)}}{\partial \textbf{x}^{(r)}}
\end{equation}

From Fig. \ref{fig:short_chain}, the partial derivative of states in short-timescale outages $\frac{\partial \textbf{x}^{(r+1)}{}'}{\partial \textbf{x}^{(r)}}$ is obtained by applying (\ref{eqn:deriv_short_state}) iteratively. And from (\ref{eqn:cost_shorttimescale}) and (\ref{eqn:deriv_short_cost}), the partial derivative of the short-timescale outage cost is derived as

\begin{equation}\label{eqn:deriv_short_sum_cost}
\frac{\partial \textbf{C}_F^{(r+1)}}{\partial \textbf{x}^{(r)}}=
\sum_{k=1}^{n_{r+1}}\frac{\partial \textbf{C}_F^{(k|r+1)}}{\partial \textbf{x}^{(r)}}
\end{equation}

\subsubsection{Re-dispatch}
Re-dispatch is usually modeled as an optimization problem. Under the DC power flow assumption, the execution of re-dispatch can be modeled as a linear programming (LP) problem\cite{yao2016risk}, which aims to minimize the distance between the actual post-dispatch state and the target state in a given time interval $\tau_D$.

\begin{equation}
\label{eqn:dispatch_exec}
\begin{aligned}
\min{}f&  =\textbf{c}_D^{\mathrm{T}}\left( \textbf{P}_{d}-\textbf{P}^*_{d} \right)+\textbf{c}_G^{\mathrm{T}}\left| \textbf{P}_{g}-\textbf{P}^*_{g} \right|\\
s.t. ~& \textbf{1}^{\mathrm{T}} \left( \textbf{P}_g-\textbf{P}_d \right) =0 \\
& -\tau_D\textbf{r}_g \leq \textbf{P}_{g}-\textbf{P}'_{g} \leq \tau_D\textbf{r}_g \\
& \textbf{P}_g^{\mathrm{min}} \leq \textbf{P}_g \leq \textbf{P}_g^{\mathrm{max}} \\
& \textbf{P}^*_d \leq \textbf{P}_d \leq \textbf{P}'_d\\
\end{aligned}
\end{equation}

\noindent where $\textbf{P}^*_{d}$ and $\textbf{P}^*_{g}$ are the target load and generation given by dispatch center, $\textbf{P}'_{d}$ and $\textbf{P}'_{g}$ are the load and generation before dispatch. $\textbf{P}_{d}$ and $\textbf{P}_{g}$ are the states to solve, i.e. the states after dispatch at time $\tau_D$. $\textbf{r}_g$ is the vector of ramping rates of all the generation buses.
(\ref{eqn:dispatch_exec}) can be briefly denoted as follows:

\begin{equation}\label{eqn:redispatch_brief}
\textbf{x}^{(r+1)}=LP_e(\textbf{p}^{(r+1)},\textbf{x}^{(r+1)}{}',\textbf{x}^{(r+1)*},\tau_D)
\end{equation}

\noindent where $\textbf{p}^{(r+1)}$ is the parameters on the $r+1$th level, such as network topology, branch parameters, branch flow limit, etc. $\textbf{x}^{(r+1)*}$ is the re-dispatch target state to fulfill, which is determined by solving another LP problem (the conventional DC-OPF in this paper) \cite{yao2016risk,dobson2001initial} as:
\begin{equation}\label{eqn:redispatch_aim_brief}
\textbf{x}^{(r+1)*}=LP_a(\textbf{p}^{(r+1)},\textbf{x}^{(r+1)}{}')
\end{equation}

From (\ref{eqn:redispatch_brief}),
$\frac{\partial \textbf{C}_R^{(r+1)}}{\partial \textbf{x}^{(r+1)}{}'}$,
$\frac{\partial \textbf{C}_R^{(r+1)}}{\partial \textbf{x}^{(r+1)*}}$,
$\frac{\partial \textbf{x}^{(r+1)}}{\partial \textbf{x}^{(r+1)}{}'}$ and
$\frac{\partial \textbf{x}^{(r+1)}}{\partial \textbf{x}^{(r+1)*}}$ can be obtained by means of Lagrange multiplier and sensitivity analysis\cite{gribik1990optimal}. Similarly, $\frac{\partial \textbf{x}^{(r+1)*}}{\partial \textbf{x}^{(r+1)}{}'}$ can be calculated from (\ref{eqn:redispatch_aim_brief}).

\subsection{Iterative calculation of terms in risk gradient}
With the analysis in III.A and the chain rule of derivatives, the terms $\frac{\partial \textbf{Pr}^{(r)}}{\partial \textbf{x}^{(0)}}$ and $\frac{\partial \textbf{C}^{(r)}}{\partial \textbf{x}^{(0)}}$ of each level $r$ can be calculated. Assume for any level $m$ ($1\leq m\leq r$), the terms $\frac{\partial \textbf{x}^{(m)}}{\partial \textbf{x}^{(0)}}$, $\frac{\partial \textbf{x}^{(m)}{}'}{\partial \textbf{x}^{(0)}}$ and $\frac{\partial \textbf{x}^{(m)*}}{\partial \textbf{x}^{(0)}}$ are obtained, then for level $r+1$ the terms $\frac{\partial \textbf{x}^{(r+1)}}{\partial \textbf{x}^{(0)}}$, $\frac{\partial \textbf{x}^{(r+1)}{}'}{\partial \textbf{x}^{(0)}}$ and $\frac{\partial \textbf{x}^{(r+1)*}}{\partial \textbf{x}^{(0)}}$ are obtained as follows
\begin{equation}\label{eqn:state_chain_prime}
\frac{\partial \textbf{x}^{(r+1)}{}'}{\partial \textbf{x}^{(0)}}=
\frac{\partial \textbf{x}^{(r+1)}{}'}{\partial \textbf{x}^{(r)}}
\frac{\partial \textbf{x}^{(r)}}{\partial \textbf{x}^{(0)}}
\end{equation}
\begin{equation}\label{eqn:state_chain_aim}
\frac{\partial \textbf{x}^{(r+1)*}}{\partial \textbf{x}^{(0)}}=
\frac{\partial \textbf{x}^{(r+1)*}}{\partial \textbf{x}^{(r+1)}{}'}
\frac{\partial \textbf{x}^{(r+1)}{}'}{\partial \textbf{x}^{(0)}}
\end{equation}
\begin{equation}\label{eqn:state_chain_state}
\frac{\partial \textbf{x}^{(r+1)}}{\partial \textbf{x}^{(0)}}=
\frac{\partial \textbf{x}^{(r+1)}}{\partial \textbf{x}^{(r+1)*}}
\frac{\partial \textbf{x}^{(r+1)*}}{\partial \textbf{x}^{(0)}}+
\frac{\partial \textbf{x}^{(r+1)}}{\partial \textbf{x}^{(r+1)}{}'}
\frac{\partial \textbf{x}^{(r+1)}{}'}{\partial \textbf{x}^{(0)}}
\end{equation}
Since $\frac{\partial \textbf{x}^{(1)*}}{\partial \textbf{x}^{(0)}}$, $\frac{\partial \textbf{x}^{(1)}{}'}{\partial \textbf{x}^{(0)}}$ and $\frac{\partial \textbf{x}^{(1)}}{\partial \textbf{x}^{(0)}}$ can be obtained from III.A, so according to (\ref{eqn:state_chain_prime})-(\ref{eqn:state_chain_state}), for any $r$ ($1\leq r \leq n$, where $n$ is the final level of cascading outage), $\frac{\partial \textbf{x}^{(r)*}}{\partial \textbf{x}^{(0)}}$, $\frac{\partial \textbf{x}^{(r)}{}'}{\partial \textbf{x}^{(0)}}$ and $\frac{\partial \textbf{x}^{(r)}}{\partial \textbf{x}^{(0)}}$ are obtained iteratively in the process of cascading outage path simulation. So $\frac{\partial \textbf{Pr}^{(r)}}{\partial \textbf{x}^{(0)}}$ and $\frac{\partial \textbf{C}^{(r)}}{\partial \textbf{x}^{(0)}}$ are calculated with
\begin{equation}\label{eqn:prob_chain}
\frac{\partial \textbf{Pr}^{(r)}}{\partial \textbf{x}^{(0)}}=
\frac{\partial \textbf{Pr}^{(r)}}{\partial \textbf{x}^{(r-1)}}
\frac{\partial \textbf{x}^{(r-1)}}{\partial \textbf{x}^{(0)}}
\end{equation}
\begin{equation}\label{eqn:cost_chain}
\resizebox{1.0\hsize}{!}{
$\displaystyle{\frac{\partial \textbf{C}^{(r)}}{\partial \textbf{x}^{(0)}} = 
\frac{\partial \textbf{C}_F^{(r)}}{\partial \textbf{x}^{(r-1)}}
\frac{\partial \textbf{x}^{(r-1)}}{\partial \textbf{x}^{(0)}}+
\frac{\partial \textbf{C}_R^{(r)}}{\partial \textbf{x}^{(r-1)}{}'}
\frac{\partial \textbf{x}^{(r-1)}{}'}{\partial \textbf{x}^{(0)}}+
\frac{\partial \textbf{C}_R^{(r)}}{\partial \textbf{x}^{(r-1)*}}
\frac{\partial \textbf{x}^{(r-1)*}}{\partial \textbf{x}^{(0)}}}$
}
\end{equation}

\subsection{Recursive form of risk gradient}
Define equivalent cascading outage cost $C'(i_{k_1}\cdots i_{k_r})$ as
\begin{equation}\label{eqn:c_prime}
\begin{aligned}
& \resizebox{1.0\hsize}{!}{$\displaystyle{C'(i_{k_1}\cdots i_{k_r})\triangleq C(i_{k_1}\cdots i_{k_r})+\sum_{i_{k_{r+1}}}\mathrm{Pr}(i_{k_{r+1}}|i_{k_1}\cdots i_{k_r})C(i_{k_1}\cdots i_{k_{r+1}})+\cdots}$}\\
& \resizebox{0.85\hsize}{!}{$\displaystyle{= C(i_{k_1}\cdots i_{k_r})+
\sum_{i_{k_{r+1}}}\mathrm{Pr}(i_{k_{r+1}}|i_{k_1}\cdots i_{k_r})C'(i_{k_1}\cdots i_{k_{r+1}})}$}
\end{aligned}
\end{equation}

(\ref{eqn:c_prime}) shows a recursive relationship, so $C'(i_{k_1}\cdots i_{k_r})$ could be calculated and updated reversely from the terminal back to the root of the MT. Similarly, define

\begin{equation}\label{eqn:r_prime}
\resizebox{0.89\hsize}{!}{$\displaystyle{R'(i_{k_1}\cdots i_{k_{r+1}})\triangleq\sum_{i_{k_{r+1}}}\mathrm{Pr}(i_{k_{r+1}}|i_{k_1}\cdots i_{k_r})C'(i_{k_1}\cdots i_{k_{r+1}})}$}
\end{equation}

With given $i_{k_1}\cdots i_{k_r}$, abbreviate all $R'(i_{k_1}\cdots i_{k_{r+1}})$ as vector $\textbf{R}^{(r)}{}'$, and $C'(i_{k_1}\cdots i_{k_r})$ as $\textbf{C}^{(r)}{}'$, then 

\begin{equation}\label{eqn:r_prime_partial}
\resizebox{0.89\hsize}{!}{$\displaystyle{\frac{\partial R'(i_{k_1}\cdots i_{k_r})}{\partial \textbf{x}^{(0)}}=
{\textbf{C}^{(r)}{}'}^{\mathrm{T}}\frac{\partial \textbf{Pr}^{(r)}}{\partial \textbf{x}^{(0)}}+
{\textbf{Pr}^{(r)}}^{\mathrm{T}} \left( \frac{\partial \textbf{C}^{(r)}}{\partial \textbf{x}^{(0)}}+
\frac{\partial \textbf{R}^{(r)}{}'}{\partial \textbf{x}^{(0)}} \right)}$}
\end{equation}
(\ref{eqn:r_prime_partial}) is also a recursive form. Note that $R^{(0)}{}'= R'$, so the gradient of risk can be computed with forward-backward scheme in the risk assessment based on MT search. 

\subsection{Forward-backward scheme of risk gradient calculation}
With (\ref{eqn:state_chain_prime})-(\ref{eqn:cost_chain}), the partial derivatives are calculated in the process of forward searching with the risk assessment \cite{yao2016risk}, as demonstrated in Algorithm 1.

\begin{table}[H]
  \centering
  \label{tab:alg1}
    \begin{tabularx}{\linewidth}{X}
      \toprule[1.5pt]
      \textbf{Algorithm 1.} Forward calculation of partial derivatives \\\midrule[1pt]
      \textbf{Step 1.} Initialize level on MT $r=0$.\\
      \textbf{Step 2.} Sample mid-timescale outages and calculate
      $\frac{\partial \textbf{Pr}^{(r+1)}}{\partial \textbf{x}^{(r)}}$ with (\ref{eqn:part_prob_state}).\\
      \textbf{Step 3.} Simulate short-timescale outages \cite{yao2016risk}, and calculate
      $\frac{\partial \textbf{x}^{(r+1)}{}'}{\partial \textbf{x}^{(r)}}$,
      $\frac{\partial \textbf{C}_F^{(r+1)}}{\partial \textbf{x}^{(r)}}$ with (\ref{eqn:cost_shorttimescale})-(\ref{eqn:deriv_short_sum_cost}).\\
      \textbf{Step 4.} Simulate re-dispatch. Calculate
      $\frac{\partial \textbf{Pr}^{(r+1)}}{\partial \textbf{x}^{(0)}}$,
      $\frac{\partial \textbf{x}^{(r+1)}}{\partial \textbf{x}^{(0)}}$ and
      $\frac{\partial \textbf{C}^{(r+1)}}{\partial \textbf{x}^{(0)}}$.\\
      \textbf{Step 5.} If the cascading outage path ends, exit and start simulation of a new path. Otherwise assign $r=r+1$ and jump to Step 2.\\
      \bottomrule[1.5pt]
    \end{tabularx}
\end{table}

After searching a cascading outage path on the MT, the risk gradient is updated reversely, as Algorithm 2 shows.

\begin{table}[H]
  \centering
  \label{tab:alg2}
    \begin{tabularx}{\linewidth}{X}
      \toprule[1.5pt]
      \textbf{Algorithm 2.} Backward update of risk gradient. \\\midrule[1pt]
      Assume the cascade path is $i_{k_1}\cdots i_{k_n}$, the variables with superscript $(r)$ correspond to $(i_{k_1}\cdots i_{k_r})$, and (0) correspond to the root state on MT. \\\midrule[0.5pt]
      \textbf{Step 1.} Define $b^{(r)}, r=1\cdots n$. If state $(i_{k_1}\cdots i_{k_r})$ has been reached before, then $b^{(r)}=0$, otherwise $b^{(r)}=1$. \\
      \textbf{Step 2.}  Assign $r=n$. Assign $\textbf{S}^{(r)}=b^{(r)}\frac{\partial C^{(r)}}{\partial \textbf{x}^{(0)}}$,
      $\Delta C^{(r)}=b^{(r)} C^{(r)}$.\\
      \textbf{Step 3.} Reverse to the previous state on the cascade path. Assign $r=r-1$.\\
      \textbf{Step 4.} Let
      $\textbf{S}^{(r)}=\mathrm{Pr}^{(r+1)}\textbf{S}^{(r+1)}+\Delta C^{(r+1)}\frac{\partial \mathrm{Pr}^{(r+1)}}{\partial \textbf{x}^{(0)}}$.\\
      \textbf{Step 5.} Let
      $\Delta C^{(r)}=b^{(r)}C^{(r)}+b^{(r+1)}\mathrm{Pr}^{(r+1)}\Delta C^{(r+1)}$.\\
      \textbf{Step 6.} If $r=0$ then exit. Otherwise jump to Step 3.\\
      \bottomrule[1.5pt]
    \end{tabularx}
\end{table}
\begin{figure}[htb]
	\centering
	\includegraphics[clip=true,scale=0.13]{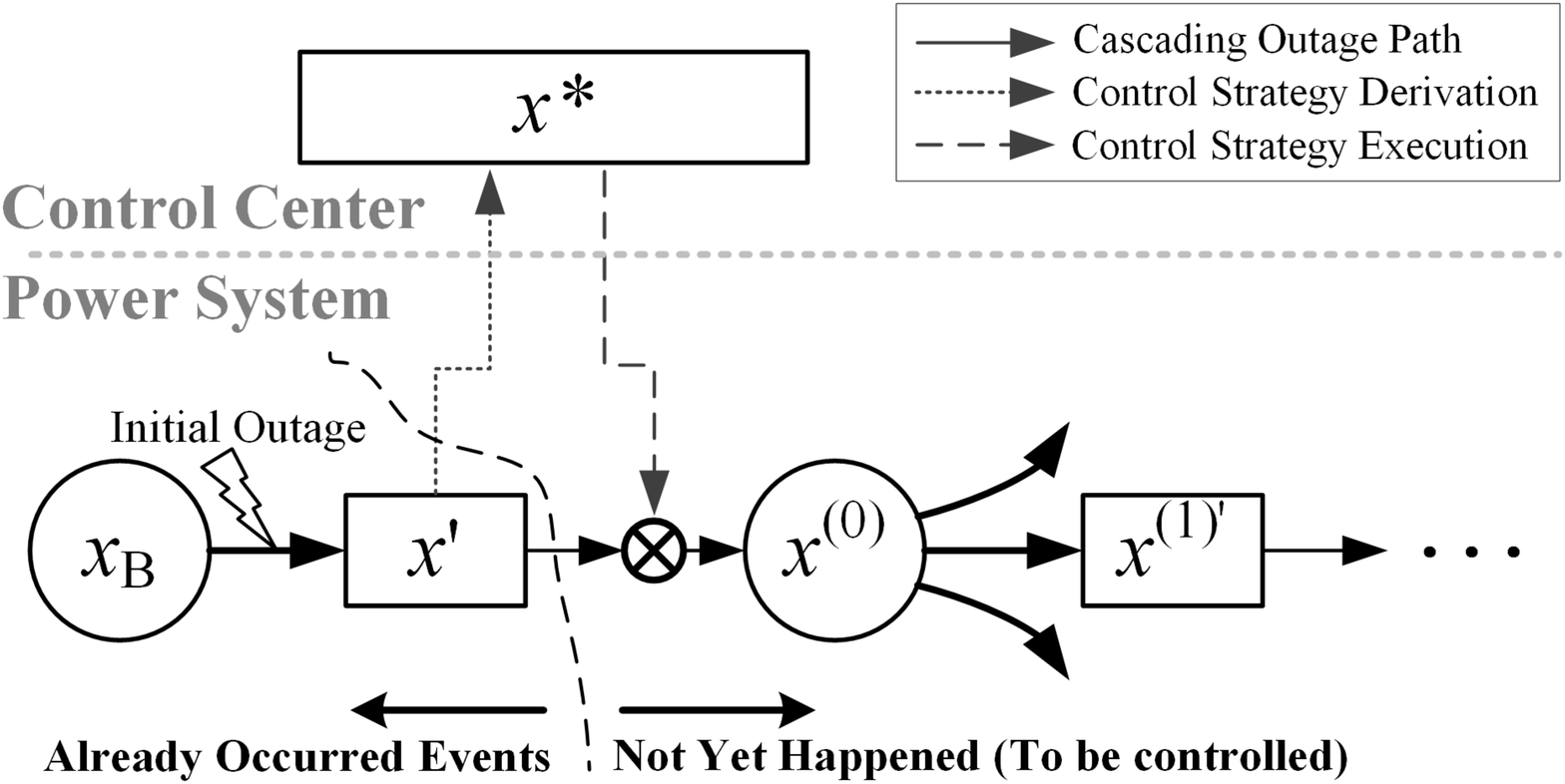}\\
	\caption{Time series of the derivation and execution of control strategy}\label{fig:timeseries}
\end{figure}

By applying Algorithms 1 and 2 repeatedly along with the forward searching and backward updating procedure of risk assessment\cite{yao2016risk}, the $\textbf{S}^{(0)}$ will converge to $\frac{\partial R^{(0)}{}'}{\partial \textbf{x}^{(0)}}$.
Note that in operations, $\textbf{x}^{(0)}$ is indirectly changed by altering the dispatch target state $\textbf{x}^*$, as Fig. \ref{fig:timeseries} shows. The gradient of risk in the space of control variables is  

\begin{equation}\label{eqn:gradient}
{\bf \Gamma}=\textbf{S}^{(0)} \frac{\partial \textbf{x}^{(0)}}{\partial \textbf{x}^*}
\end{equation}

\section{Implementation of Risk Management}
\subsection{Full optimization model of risk management (RM)}
After obtaining the risk gradient in the space of control variables, the risk management (RM) optimization model is established based on the generic form of (\ref{eqn:mitigation_linearization}) as
\begin{equation}
\label{eqn:crm}
\begin{aligned}
\min{}f & =-\textbf{c}_D^{\mathrm{T}}\left( \textbf{P}^*_{d}-\textbf{P}_{d} \right)+\textbf{c}_G^{\mathrm{T}}\left| \textbf{P}^*_{g}-\textbf{P}_{g} \right|\\
s.t.~ & -{\bf \Gamma}\cdot \left[
\begin{aligned}
& \textbf{P}^*_d-\textbf{P}^*_{d0}\\
& \textbf{P}^*_g-\textbf{P}^*_{g0}\\
\end{aligned} \right]
\leq R_E-R'(\textbf{x}^{(0)*}_0) \\
& \textbf{1}^{\mathrm{T}} \left( \textbf{P}^*_g-\textbf{P}^*_d \right) =0 \\
& -\textbf{F}^{\mathrm{max}}\leq \textbf{y}_D\textbf{MY}^+ \left( \textbf{P}^*_g-\textbf{P}^*_d \right) \leq \textbf{F}^{\mathrm{max}} \\
& \textbf{P}_g^{\mathrm{min}} \leq \textbf{P}^*_g \leq \textbf{P}_g^{\mathrm{max}} \\
& \textbf{0} \leq \textbf{P}^*_d \leq \textbf{P}_d\\
\end{aligned}
\end{equation}
\noindent where the first constraint is the risk constraint using the risk gradient. $\textbf{x}^{(0)*}_0=[{\textbf{P}^*_{d0}}^{\mathrm{T}},{\textbf{P}^*_{g0}}^{\mathrm{T}}]^{\mathrm{T}}$ is the target state from the original dispatch strategy, $R'(\textbf{x}^{(0)*}_0)$ is the risk of subsequent cascades of original dispatch strategy, and $R_E$ is the expected risk after the RM. The other constraints are the limits of branches, generators and loads. Here the variables are continuous, so the RM is an LP problem. If there are discrete variables, then RM will become a mixed-integer linear programming (MILP) problem.

The (\ref{eqn:crm}) reduces cascading outages risk by setting its solution as the target state of re-dispatch. The extent of reduction of cascading outage risk is adjusted by changing the expected subsequent cascade risk $R_E$. The constraint will force the solution to a less risky state and the control cost is expected to be higher. Denote the expected risk decrease as $\Delta R=R'(\textbf{x}^{(0)*}_0)-R_E \geq 0$, then the bigger $\Delta R$, the more reduction of risk is expected.

\subsection{Iterative risk management (IRM)}
\begin{figure}[htb]
	\centering
	\includegraphics[clip=true,scale=0.75]{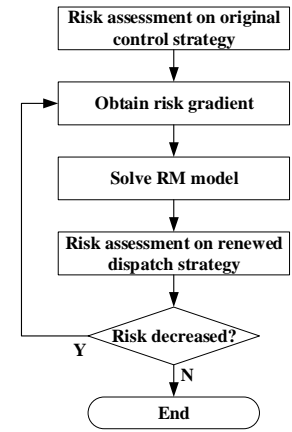}\\
	\caption{The procedure of IRM}\label{fig:ICRM}
\end{figure}
The RM model is based on the linearization of risk at the original operating point. When $\Delta R$ goes outside an effective region of linearization, then there will be considerable linearization errors. To overcome such limitation, consider iterating the procedure of RM so as to accumulate the effect of linearization-based RM step by step. The procedure of iterative RM (IRM) is shown as Fig. \ref{fig:ICRM}.

The IRM first assesses risk on the original control strategy, and solves the RM problem (\ref{eqn:crm}). The new dispatch strategy is then evaluated with risk assessment. If the risk is decreased, then the strategy is expected to be effective. Such a procedure is iterated until the risk does not decrease.

\subsection{Framework of RM/IRM application}
\begin{figure}[htb]
  \centering
  \includegraphics[clip=true,scale=0.7]{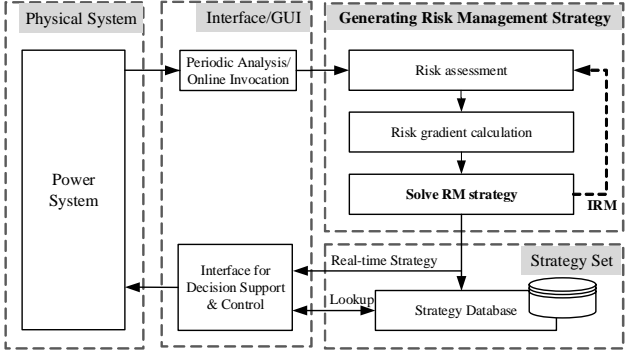}\\
  \caption{The framework of RM/IRM application}\label{fig:CRM_application}
\end{figure}
The RM and IRM can be used off-line to generate control strategies on given set of system working conditions. The generated strategies are stored in a database and can be extracted when corresponding events occur. Moreover, the RM and IRM also have potential of online assessment and decision support. The framework of RM/IRM application is demonstrated in Fig. \ref{fig:CRM_application}.

\section{Case Studies}

\subsection{Convergence of risk gradient in RTS-96 system}
According to III.C, the calculation of risk gradient can be integrated into the forward-backward scheme of risk assessment. To efficiently derive risk management strategy, the convergence of risk gradient is of great significance.

In the computation process, the risk gradient after the $m$th search attempt is denoted as ${\bf \Gamma}_m$, and the converged risk gradient is denoted as ${\bf \Gamma}^*$. To evaluate the convergence profile of risk gradient, propose the following convergence indices:
\begin{equation}\label{eqn:deltam}
\delta_m=\frac{\|{\bf \Gamma}_m-{\bf \Gamma}^*\|}{\|{\bf \Gamma}_1-{\bf \Gamma}^*\|}
\end{equation}
\begin{equation}\label{eqn:deltamto1}
\delta_m^{\mathrm{dir}}=\left\|\frac{{\bf \Gamma}_m}{\|{\bf \Gamma}_m\|}-\frac{{\bf \Gamma}^*}{\|{\bf \Gamma}^*\|}\right\|
\end{equation}

\begin{figure}[htb]
  \centering
  \includegraphics[clip=true,scale=0.23]{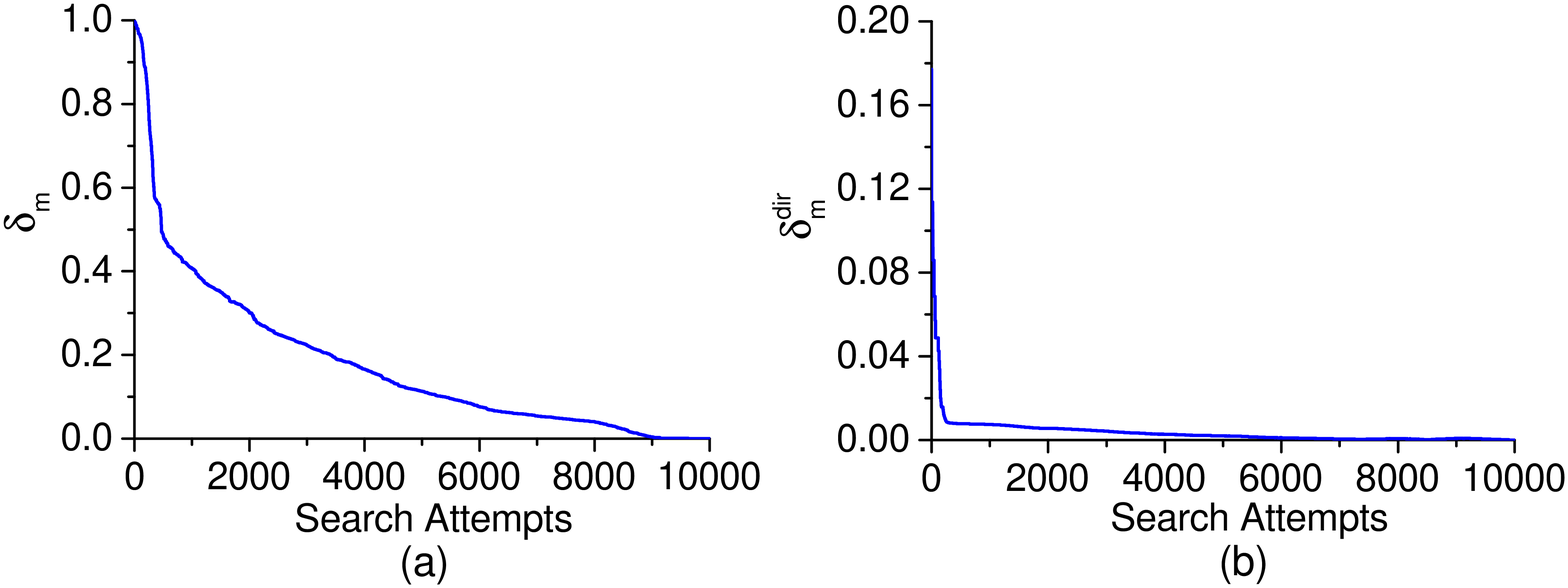}\\
  \caption{Convergence of (a) risk gradient (b) the direction of risk gradient}\label{fig:rts_crm_d}
\end{figure}

$\delta_m$ reflects the convergence of the vector of risk gradient, and $\delta_m^{\mathrm{dir}}$ evaluates the convergence of the direction of risk gradient. The test is conducted on the RTS-96 3-area system, in which the parameters are set as $T_{\max{}}=150\mathrm{min}$, $\tau_D=15\mathrm{min}$. After 10000 times of tree search, the risk, $\delta_m$ and $\delta_m^{\mathrm{dir}}$ are all considered as converged. Fig. \ref{fig:rts_crm_d} demonstrates the convergence profile of risk gradient.

\begin{figure}[htb]
	\centering
	\includegraphics[clip=true,scale=0.25]{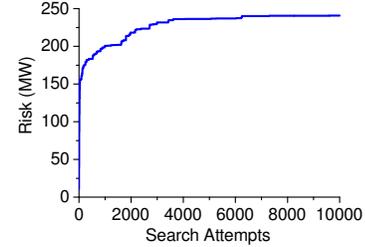}\\
	\caption{Convergence profile of risk \cite{yao2016risk}}\label{fig:rts_crm_r}
\end{figure}

The convergence of risk gradient is slower than that of risk \cite{yao2016risk} (Fig. \ref{fig:rts_crm_r}). This is caused by the more complicated form of risk gradient, so the derivation of risk gradient may cost more computation time than risk assessment. However, the convergence of the direction of risk gradient only requires several hundreds of search attempts, which is much faster than the convergence of the vector of risk gradient. Actually, only obtaining the direction of risk gradient is enough for risk reduction, so in situations that require fast computation, the number of tree search attempts can be significantly reduced. Nevertheless, in such a case, the accuracy of estimating the extent of risk reduction will be lower.

\subsection{Effectiveness of RM}
\subsubsection{RTS-96 System}
\begin{figure}[htb]
  \centering
  \includegraphics[clip=true,scale=0.30]{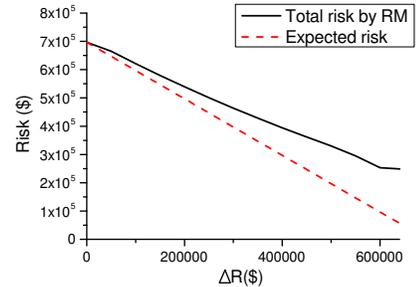}\\
  \caption{Convergence profile of risk under RM}\label{fig:rts_crm}
\end{figure}

After validating the accuracy and the convergence of the calculation of risk gradient, the effectiveness of RM in the RTS-96 3-area system model is tested. Here all the cost and risk are converted into economic metrics. Assume that adjusting 1MW of generation in an interval $\tau_D$ costs \$100, and 1MW load loss in an interval corresponds to the loss of \$10000. Set the initial failure on branches 22, 23 and 24. When utilizing conventional re-dispatch, the total risk (i.e. the cost of control plus the risk of subsequent cascading outages) is \$696775. Then use the RM to reduce cascading outage risk and evaluate the performance with risk assessment.
The effect of RM under different values of $\Delta R$ is shown in Fig. \ref{fig:rts_crm}. The RM effectively decreases risk within a certain range, but the risk stops decreasing when $\Delta R$ reaches around \$600000, which means the linearization is no longer effective.

\begin{figure}[htb]
  \centering
  \includegraphics[clip=true,scale=0.33]{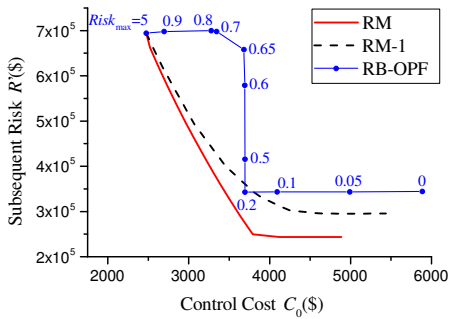}\\
  \caption{Risk reduction performance of RM, RM-1 and RB-OPF}\label{fig:rts_tradeoff}
\end{figure}

Next, the performance of RM with several existing methods are compared. The risk-based OPF method (RB-OPF)\cite{xiao2007risk} and a variant of ref. \cite{capitanescu2015enhanced} are tested in this case. In the RB-OPF, the risk can be reduced by lowering the upper limit of severity-based risk index $Risk_{\max{}}$. Decreasing the $Risk_{\max{}}$ is expected to reduce the risk. Ref. \cite{capitanescu2015enhanced} changed the risk constraint of RB-OPF/SCOPF as the expected load shedding after N-1 outages. Since \cite{capitanescu2015enhanced} is similar to the risk management considering only the first level of cascading outages, the risk assessment and calculating the risk gradient is conducted only to level-1 outages, as a variant of \cite{capitanescu2015enhanced}. Such a variant of \cite{capitanescu2015enhanced} is named as RM-1. Fig. \ref{fig:rts_tradeoff} compares the costs of control and subsequent cascading outage risks under the RM, RM-1 and RB-OPF.
The RM achieves a lower risk than the other two approaches at the same level of cost, so it is verified that the RM is more effective in reducing the cascading outage risk. 

The RM-1, regarded as an simplified version of RM, does not consider the multi-level cascading outages, so the accuracy of reducing the cascading outage risk is not as satisfactory as RM. It should be noted that although the result of RM-1 in Fig. \ref{fig:rts_tradeoff} still seems fine, the effectiveness of RM-1 actually highly depends on the distribution of the risk on different levels of cascading outages. In this test case, the risk mainly exists on the level-1 outages, so the RM-1 does not have significant difference from RM. But if substantial risk is on other levels of outages, the accuracy of RM-1 will be significantly affected.

Fig. \ref{fig:rts_tradeoff} further annotates the cost-risk relationship with different values of $Risk_{\max{}}$ in the RB-OPF. The result shows that the relationship between cascading outage risk and $Risk_{\max{}}$ is non-smooth, which will cause difficulties in managing the cascading outage risk. By comparison, the RM proposed in this paper achieves a better performance in reducing risk, and the amount of risk reduction is more controllable. 

\begin{figure}[htb]
	\centering
	\includegraphics[clip=true,scale=0.35]{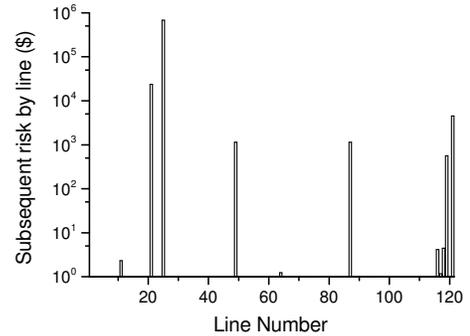}\\
	\caption{Risk of subsequent cascading outages by level-1 line outage.}\label{fig:rts_riskByLine}
\end{figure}

\begin{figure}[htb]
	\centering
	\includegraphics[clip=true,scale=0.21]{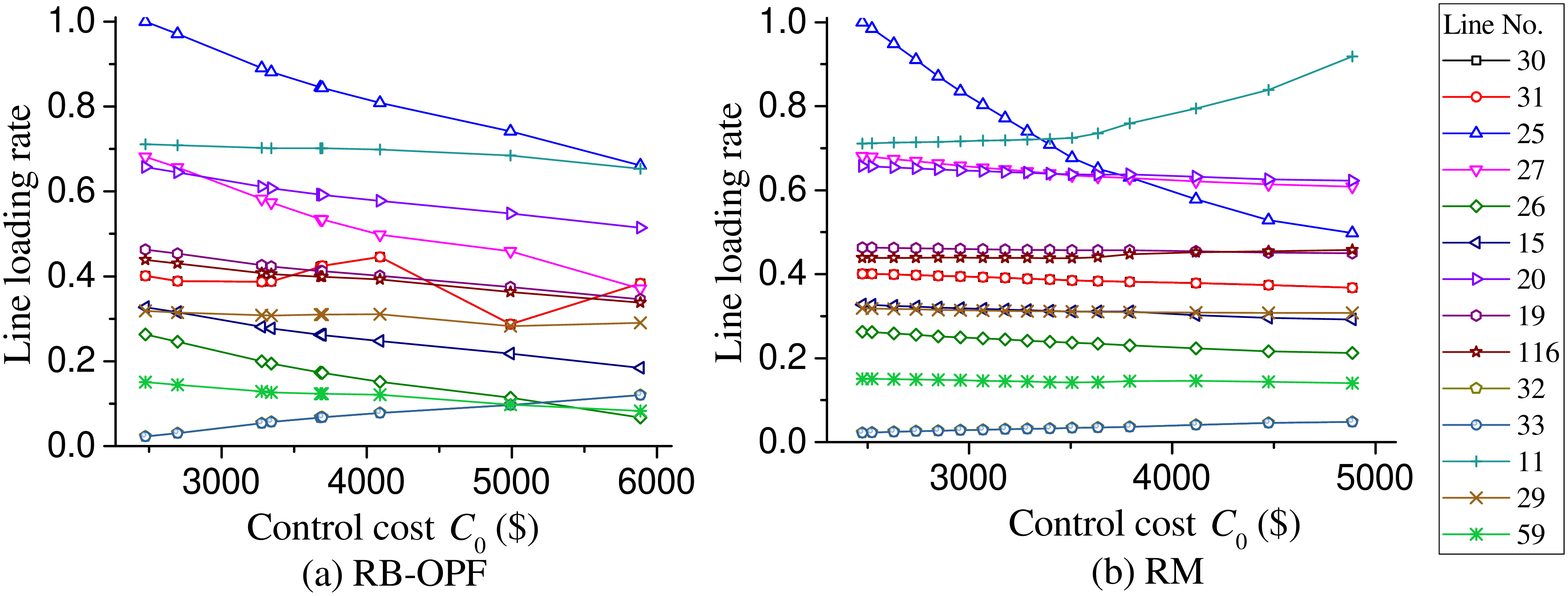}\\
	\caption{Loading rates of selected lines under (a) RB-OPF, (b) RM.}\label{fig:rts_loading_merge}
\end{figure}

\begin{figure}[htb]
	\centering
	\includegraphics[clip=true,scale=0.35]{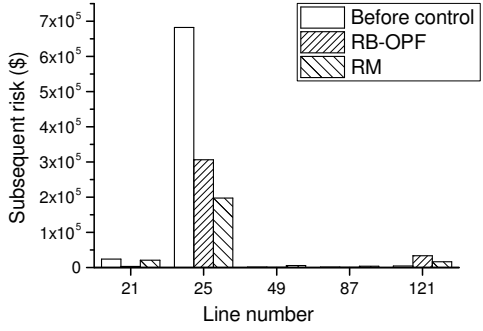}\\
	\caption{Cascading outage risks of selected level-1 outages.}\label{fig:rts_risk_postcontrol}
\end{figure}

The mechanisms of RB-OPF and RM are further analyzed and compared. In RB-OPF, as $Risk_{\max{}}$ reduces, the RB-OPF optimization model adjusts the loading of lines to reduce the severity of post-outage states of all the lines in the system. However, as RB-OPF does not have accurate knowledge of the most risky cascading outage patterns, it is not as effective as RM for cascading outages. The risk in RB-OPF is described as the outage probability times post-contingency severity, but the severity does not very well reflect the actual loss (load or economic loss) of the potential subsequent cascading outages, so the RB-OPF tends to lower the loading of lines system-wide, disregarding that the risk under many subsequent outages are actually very low. As Fig. \ref{fig:rts_riskByLine} shows, the distribution of subsequent cascading outage risk in terms of level-1 line outage is very uneven: most risk is concentrated on the level-1 outages of lines 21, 25, 49, 87 and 121 (note that there are only 120 lines in RTS-96 system, and the "line 121" stands for no outage in the studied time interval). Particularly, the risk after line 25 outage accounts for about 95\% of the total risk. We select the lines whose loading rates are significantly changed by these methods, and the relationship between line loading rates and control cost. As shown in Fig. \ref{fig:rts_loading_merge}(a), the RB-OPF reduces the risk by suppressing the loading level on many lines even if the actual risk of the following cascading outages is relatively trivial. In contrast, Fig. \ref{fig:rts_loading_merge}(b) shows the line loading rates under RM. Since the risk assessment identifies that the outage of line 25 is the most risky, the RM mainly reduces the loading of line 25, and thus more effectively reduces the overall subsequent cascading outage risk with lower control cost. Fig. \ref{fig:rts_risk_postcontrol} compares the distribution of risk before control, after RB-OPF and after RM. The results show that RM identifies the "bottleneck" of cascading outage risk and better concentrates the control effort on the risky cascading outage patterns. As a result, the overall risk under RM is lower at the same level of control cost. This is attributed to the feature of the RM optimization model that makes use of the information of risk distribution obtained from the risk assessment and risk gradient.

%

\subsubsection{US-Canada Northeast system}
\begin{figure}[htb]
	\centering
	\includegraphics[clip=true,scale=0.40]{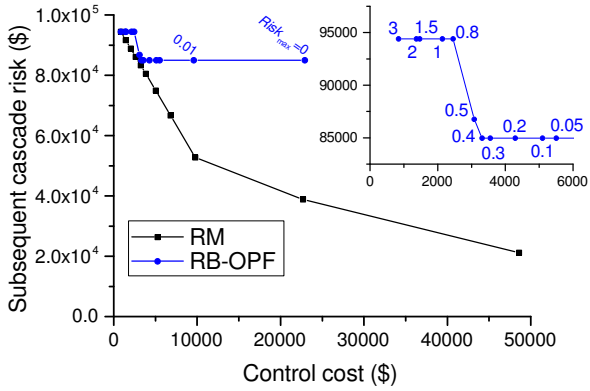}\\
	\caption{Risk reduction by RM and RB-OPF in US-Canada northeast system}\label{fig:npcc_rm_comparison}
\end{figure}

Next, the RM is tested on a real-world system model, i.e. the US-Canada Northeast power grid model \cite{7254205} retaining all generator buses and the buses at 230 kV or above. The system contains 410 buses (287 load buses and 233 generator buses, note that a bus may be both load bus and generator bus) and 882 branches.
In this case, increase the system load to 1.5 times of the base load and set initial outages on branches 320 and 321. Fig. \ref{fig:npcc_rm_comparison} compares the cost-risk profiles obtained by RM and RB-OPF, respectively. It can be seen that the RM reduces the risk by over 70\%, while the RB-OPF only reduces the risk by about 10\%. The effect of RB-OPF is limited due to its mechanism. According to the definition of severity function in RB-OPF, if the loading rates of a line under all N-1 states are below 90\%, then the risk of the line is 0. So when setting the $Risk_{\max{}}=0$, the RB-OPF will guarantee the loading rates of all the lines under all N-1 states under 90\%. However, in this case, substantial risk may still exist under some cascading outage patterns, which cannot be reduced by RB-OPF. In contrast, since RM uses the information obtained from the risk assessment of cascading outages, it more effectively reduces the risk of subsequent cascading outages.

The computational efficiency of RB-OPF and RM is also tested. Table \ref{tab:rm_npcc} shows the average time of RM and RB-OPF in US-Canada northeast system. The methods are developed with MATLAB and are tested on a PC with 2.6GHz CPU and 32GB RAM. Similar with the computational strategies proposed by \cite{wang2013risk}, the RB-OPF is realized with the technique of benders decomposition to handle large-system case. Since the RB-OPF considers the severity function of all the lines under all the N-1 outages, the scale of the optimization problem grows in proportion to the square of the number of branches in the system. In the US-Canada northeast system with 882 branches, the average computation time of RB-OPF is 72.29s. While the optimization model of RM has much smaller scale than RB-OPF, and the average computation time is less than 1s. However, since the optimization model of RM requires risk gradient, the computation time of risk assessment and risk gradient calculation should also be included. In this case, each risk assessment uses 200 Markovian tree search attempts, and the average computation time is longer than that of RB-OPF.

\begin{table}[htb]
	\centering
	\caption{Time consumption comparison in US-Canada Northeast system}
	\label{tab:rm_npcc}
	\begin{tabularx}{\linewidth}{p{3.3cm}@{}*2{>{\centering\arraybackslash}X}@{}}
		\toprule[1.5pt]
		\multirow{2}*{\textbf{Subprograms}}  & \multicolumn{2}{c}{\textbf{Time consumption (s)}} \\
		& RM & RB-OPF \\\midrule[1pt]
		Risk assessment  & 79.41 & --\\
		Computation of risk gradient & 25.40 & -- \\
		Solving RM/RB-OPF model & 0.36 & 72.29  \\ \midrule[0.5pt]
		\textbf{Total}  & \textbf{105.17} & \textbf{72.29}     \\
		\bottomrule[1.5pt]
	\end{tabularx}
\end{table}

\subsection{Risk-cost coordination realized by IRM}
\subsubsection{RTS-96 System}
\begin{figure}[htb]
  \centering
  \includegraphics[clip=true,scale=0.35]{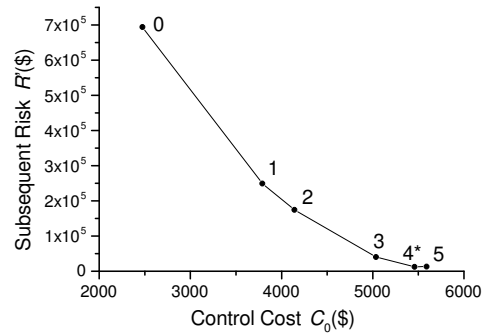}\\
  \caption{Cost-risk trajectory of IRM in RTS-96 test system}
  \label{fig:rts_icrm}
\end{figure}

The performance of the RM is limited only within a range where linearization is effective. Fig. \ref{fig:rts_crm} shows that with the RM, the risk stops decreasing at around $\$2.5\times 10^5$. IRM keeps updating risk gradient at new operating points and the risk is further reduced. Table \ref{tab:CRM_iter_rts} and Fig. \ref{fig:rts_icrm} demonstrates the cost-risk relationship derived by IRM.
\begin{table}[htb]
  \centering
  \caption{Cost-risk in the iteration process in RTS-96 test system}
  \label{tab:CRM_iter_rts}
    \begin{tabularx}{\linewidth}{p{0.1cm}@{}*5{>{\centering\arraybackslash}X}@{}}
      \toprule[1.5pt]
      & Round  & $\Delta R$ (\$)    & Control cost(\$)& Subsequent risk(\$)    & Total risk(\$)\\\midrule[1pt]
      & 0      & --                 & 2475            & 694300                 & 696775        \\
      & 1      & 600000             & 3789.8          & 249590                 & 253379.8      \\
      & 2      & 100000             & 4141.6          & 174840                 & 178981.6      \\
      & 3      & 100000             & 5034.9          & 40738                  & 45772.9       \\
      & 4      & 20000              & 5456.4          & 12434                  & 17890.4       \\
      & 5      & 1000               & 5590            & 13132                  & 18722         \\
      \bottomrule[1.5pt]
    \end{tabularx}
\end{table}

The results indicate that the IRM can effectively overcome the limitation of linearization with the RM and further reduce risk of cascading outages. After 4 rounds of iterations, the subsequent risk of cascading outages has reduced by 97.3\%.

\subsubsection{US-Canada Northeast system}
\begin{figure}[htb]
  \centering
  \includegraphics[clip=true,scale=0.35]{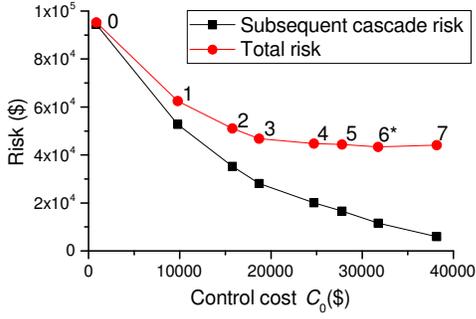}\\
  \caption{Cost-risk trajectory of IRM in US-Canada Northeast system}
  \label{fig:npcc_icrm}
\end{figure}

Next, the IRM is tested on the US-Canada Northeast power grid model. The cost-risk characteristics derived by IRM are demonstrated in Fig. \ref{fig:npcc_icrm}.

The results indicate that the total risk at the 6th iteration reaches the lowest among all iterations, where the total risk is expected to decrease by 54.5\%, and the subsequent cascade risk drops significantly by 93.6\%. The subsequent cascade risk is even lower in the 7th iteration, but the drop in subsequent cascade risk is offset by a substantially high cost, which causes a higher total risk than that of the 6th iteration. In practice, the adopted strategy for risk management may vary depending on the risk preference. Therefore, the strategies in Fig. \ref{fig:npcc_icrm} can also be regarded as the results of multi-objective optimization of risk and cost, and the control strategy for risk management can be selected depending on the risk preference.

Regarding the computational efficiency, the average performance of a single IRM run in the US-Canada Northeast system model is shown in Table \ref{tab:icrm_npcc}. The result shows that solving the RM models takes only several seconds, but much more time is consumed in cascading outage simulation, risk assessment and calculating the risk gradient. The speed of computation can be further enhanced with parallel computing on high-performance computation platform, and this method also has potentials for on-line application on a period of 5-15 minutes for operators' decision support to prevent cascading outages.
\begin{table}[htb]
  \centering
  \caption{Time consumption of IRM in US-Canada Northeast system}
  \label{tab:icrm_npcc}
    \begin{tabularx}{\linewidth}{p{5.4cm}X}
      \toprule[1.5pt]
      \textbf{Subprograms of IRM}  & \textbf{Time consumption (s)} \\\midrule[1pt]
      Cascading outage simulation \& risk assessment  & 660.41\\
      Computation of risk gradient & 237.42 \\
      Solving RM model & 3.48  \\
      \textbf{Total}  & \textbf{901.31}     \\
      \bottomrule[1.5pt]
    \end{tabularx}
\end{table}

As for the drawbacks of the proposed approach, it is observed that the accuracy of risk gradient calculation may decrease as the size of the power system grows. This is because as the system size grows, the number of nonlinear behaviors (e.g. switched active constraints in dispatch model, etc.) also grows. To maintain desired accuracy, more iterations in IRM may be necessary, and the computation speed will be adversely affected. The estimation of effective linearization region and the derivation of desirable step size in IRM will be studied in the future. Moreover, since the calculation of risk gradient requires to store all the sensitivity matrices $\frac{\partial \textbf{x}^{(r)*}}{\partial \textbf{x}^{(0)}}$, $\frac{\partial \textbf{x}^{(r)}{}'}{\partial \textbf{x}^{(0)}}$ and $\frac{\partial \textbf{x}^{(r)}}{\partial \textbf{x}^{(0)}}$, the memory usage in large-scale systems will be high. This problem can be alleviated by using compressed storage (CS) of the sensitivity matrices, since most elements in the sensitivity matrices have very low absolute values (in large systems, generally less than 1\% of the elements have absolute values larger than $10^{-3}$, and less than 10\% have absolute value larger than $10^{-5}$). Compressing the matrices by dropping low-absolute-value elements practically does not affect the accuracy, but can significantly decrease memory usage.

\subsubsection{1354-bus Mid-European System \cite{josz2016ac}}
\begin{figure}[htb]
	\centering
	\includegraphics[clip=true,scale=0.35]{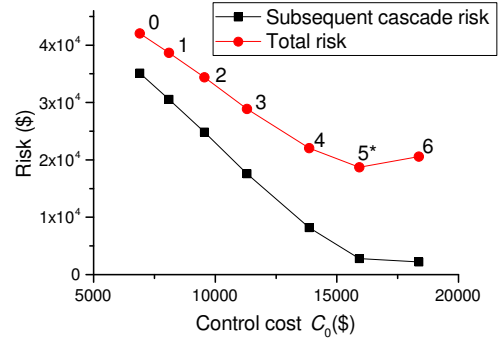}\\
	\caption{Cost-risk trajectory of IRM in Mid-European system}
	\label{fig:mideu_icrm}
\end{figure}

To further demonstrate the performance, the proposed approach is tested on a larger Mid-European backbone system model that appears in \cite{josz2016ac,zimmerman2011matpower}. The system has 1354 buses (including 260 generation buses and 688 load buses) and 1991 branches. The branch flow limits were modified to secure N-1. We maintain the same number of MT search attempts as the US-Canada Northeast System case, and run the IRM. The effectiveness of IRM is shown in Fig. \ref{fig:mideu_icrm}, and the computational time is shown in Table \ref{tab:icrm_mideu}. The computation can be finished within 1 hour with our desktop computer, which shows promising potential for online decision support in practical systems. Also, the speed can be further enhanced with parallel computing on high-performance platforms (currently the approach is developed and tested on Matlab and has not been further optimized for performance).

\begin{table}[htb]
	\centering
	\caption{Time consumption of IRM in Mid-European system}
	\label{tab:icrm_mideu}
	\begin{tabularx}{\linewidth}{p{5.4cm}X}
		\toprule[1.5pt]
		\textbf{Subprograms of IRM}  & \textbf{Time consumption (s)} \\\midrule[1pt]
		Cascading outage simulation \& risk assessment  & 1806.17\\
		Computation of risk gradient & 1449.69 \\
		Solving RM model & 6.079  \\
		\textbf{Total}  & \textbf{3261.94}     \\
		\bottomrule[1.5pt]
	\end{tabularx}
\end{table}

For large scale systems, the CS can be adopted to avoid memory overflow. Table \ref{tab:memory} compares the performance with and without CS. The threshold of absolute sensitivity value is set as $10^{-5}$. Although the CS causes some computational overhead in compressing and indexing matrices, the memory usage is significantly reduced. Per the observation from the test results on more systems, the computation time complexity when using the CS is approximately $O(N^2)$, where $N$ is the number of buses. Moreover, it is observed that the number of remaining sensitivity elements after CS grows approximately linearly with system size, while the full sensitivity matrix grows quadratically with system size. Such a desirable spatial complexity of CS contributes to the less overall computation time in large-scale systems cases, as Table \ref{tab:memory} shows. Therefore, the CS is recommended for large systems.

\begin{table}[htb]
	\centering
	\caption{Time \& Memory usage of risk gradient computation}
	\label{tab:memory}
	\begin{tabularx}{\linewidth}{p{1.7cm}@{}*4{>{\centering\arraybackslash}X}@{}}
		\toprule[1.5pt]
		\multirow{2}*{\textbf{Systems}}  & \multicolumn{2}{c}{\textbf{Without CS}} & \multicolumn{2}{c}{\textbf{With CS}} \\
		& Time(s) & Memory(GB) & Time(s) & Memory(GB) \\\midrule[1pt]
		US-Canada    & 237.42  & 4.87 & 670.10  & 0.76 \\
		Mid-European & 2184.60 & 30.25& 1449.69 & 3.32 \\
		\bottomrule[1.5pt]
	\end{tabularx}
\end{table}

\section{Conclusion}
This paper proposes an approach for risk management of cascading outages based on risk gradient and MT search. The expansion of risk corresponding to the MT structure is linearized in the space of control variables and then the gradient of risk is obtained. The computation of risk gradient adopts an efficient forward-backward algorithm, which can be combined with the procedure of risk assessment based on Markovian tree search, and the risk gradient has good convergence profile. With the risk gradient, the constraint of risk is established and incorporated into a dispatch model to formulate the risk management (RM) optimization model. The RM minimizes control cost while limiting the cascading outage risk under a given level for a desirable trade-off between cost and risk. The risk and cost in the RM model have clear physical meanings and thus can practically give insights for operators' decision support. The effectiveness of RM is verified on the RTS-96 3-area system and a 410-bus US-Canada system model.

To overcome the limitation of linearization in the RM, the iterative RM (IRM) approach is proposed to achieve more effective reduction of cascading outages risk. The test cases on the US-Canada northeast system and 1354-bus Mid-European system verify the effectiveness of the RM and IRM. Moreover, compressed storage (CS) of sensitivity matrix technique is introduced to significantly reduce memory usage without compromising the accuracy, which further improves the practicality of the proposed approach in large-scale system applications. The RM and IRM may be utilized in on-line decision support for preventive control against potential cascading outages following possible initial contingency scenarios.



%
%

\ifCLASSOPTIONcaptionsoff
  \newpage
\fi

\bibliographystyle{IEEEtran}
\bibliography{refs/refs}

\end{document}